\documentclass[11pt]{article}
\usepackage{amsmath}
\usepackage{graphicx}
\usepackage{amssymb}
\usepackage{amsfonts}
\usepackage{latexsym}
\usepackage{setspace}
\usepackage{mathptmx}

\RequirePackage{cite}%
\renewcommand{\citeleft}{\bgroup\normalfont[}
\renewcommand{\citeright}{]\egroup}%

\newcommand{\nn}{\nonumber}
\newcommand{\tr}{{\rm tr\,}}
\newcommand{\be}{\begin{equation}}
\newcommand{\ee}{\end{equation}}
\newcommand{\ba}{\begin{eqnarray}}
\newcommand{\ea}{\end{eqnarray}}
\newcommand{\bal}{\begin{align}}
\newcommand{\eal}{\end{align}}
\newcommand{\lb}{\label}
\newcommand{\ol}{\overline}
\newcommand{\e}{{\rm e}}
\newcommand{\dd}{{\rm d}}
\newcommand{\ii}{{\rm i}}

\def\E{{\cal E}}

\begin{document}

\title{\rightline{\footnotesize LAPTH-003/11}
\vskip0.5cm {\Large\textbf{Phantom Black Holes and Sigma Models}}}

\author{Mustapha Azreg-A\"{\i}nou\textsuperscript{(a)}\thanks{E-mail: azreg@baskent.edu.tr}\\
G\'erard Cl\'ement\textsuperscript{(b)}\thanks{E-mail: gclement@lapp.in2p3.fr}\\
J\'ulio C. Fabris\textsuperscript{(c,d)}\thanks{E-mail: fabris@pq.cnpq.br}\\
Manuel E. Rodrigues\textsuperscript{(b,c)}\thanks{E-mail: esialg@gmail.com}\\\\
{\small \textsuperscript{(a)} Ba\c{s}kent University, Department of Mathematics,}\\
{\small  Ba\u{g}l\i ca Campus, Ankara, Turkey}\\
{\small \textsuperscript{(b)} Laboratoire de  Physique Th\'eorique LAPTH (CNRS),}\\
{\small B.P.110, F-74941 Annecy-le-Vieux cedex, France}\\
{\small \textsuperscript{(c)} Universidade Federal do Esp\'{\i}rito Santo,} \\
{\small Centro de Ci\^{e}ncias Exatas - Departamento de F\'{\i}sica,} \\
{\small Av. Fernando Ferrari, 514 - Campus de Goiabeiras,} \\
{\small CEP29075-910 - Vit\'{o}ria/ES, Brazil}\\
{\small \textsuperscript{(d)} IAP, Institut d'Astrophysique de Paris,} \\
{\small  98 bis, Bd Arago - 75014 Paris, France}}

\date{18 February 2011}

\maketitle

\begin{abstract}
We construct static multicenter solutions of phantom
Einstein-Maxwell-dilaton theory from null geodesics of the target
space, leading to regular black holes without spatial symmetry for
certain discrete values of the dilaton coupling constant. We also
discuss the three-dimensional gravitating sigma models obtained by
reduction of phantom Einstein-Maxwell, phantom Kaluza-Klein and
phantom Einstein-Maxwell-dilaton-axion theories. In each case, we
generate by group transformations phantom charged black hole
solutions from a neutral seed.

\vspace{3mm}

\end{abstract}

\newpage

\setcounter{equation}{0}
\section{Introduction}

Phantom gravitating field theories are theories where one or more of
the matter fields appear in the action with a kinetic term of the
``wrong sign", so that they are coupled repulsively to gravity. This
implies the violation of the null energy condition $p + \rho \geq
0$. At the quantum level, such hypothetical fields could form a
``ghost condensate" vacuum, which would lead to modifications of
gravity in the infra-red limit \cite{arkani-hamed}. Present
observational evidence cannot rule out this possiblity, and even
under certain conditions it seems to favor a phantom scenario. For
example, the seven-year WMAP results, combined with $BAO$ and $H_0$
data, indicate that, imposing a
constant equation of state parameter $\omega = p/\rho$, it comes out
that $\omega = - 1.10\pm 0.14$ for a flat universe, and $\omega = -
1.44\pm 0.27$ for a non-flat universe \cite{komatsu}. Ghost
condensate models may also lead to a singularity-free primordial
scenario generating a perturbation spectrum in agreement with the
present observations \cite{brand}.

The presence of phantom fields in the action results in interesting
new possibilities for stationary solutions, such as wormhole
solutions in the case of a phantom scalar field \cite{worm},
zero-mass black holes \cite{gira}, and cold black holes with a
multiply degenerate horizon and an infinite horizon area
\cite{cold1,cold2}. The static, spherically symmetric black hole
solutions to Einstein-Maxwell-dilaton (EMD) theory with phantom
Maxwell and/or dilaton field were systematically investigated in
\cite{phantom}. Nine classes of asymptotically flat phantom black
holes, as well as two classes of non-asymptotically flat phantom
black holes, were found, and their causal structure was analyzed,
leading to a rich variety of sixteen different types of causal
structures.

The purpose of the present work is to go beyond the static
spherically symmetric approach of \cite{phantom} by using nonlinear
sigma-model techniques. Dimensional reduction of a sector of four-
or higher-dimensional gravitating field theories to three dimensions
leads to a gravitating sigma model (see e.g. \cite{BMG,book}).
Besides providing a path (essentially equivalent to that followed in
\cite{phantom}) to the construction of static spherically symmetric
solutions as target space (or potential space) geodesics, the
sigma-model approach can be used to generate new solutions in two
ways. First, solutions without spatial symmetry, also called
multicenter, or BPS, solutions generalizing the static
Majumdar-Papapetrou solutions of Einstein-Maxwell (EM) theory
\cite{papa} and their stationary counterpart \cite{IWP} can be
constructed from null geodesics of the target space \cite{spat,bps}.
Second, in special cases the target space is a symmetric space, or
coset, leading to the possibility of generating new solutions, e.g.
stationary axisymmetric solutions, by applying group transformations
to the coset representative of a seed solution.

In the present paper, we shall follow the two approaches just
described. In the next Section, we consider the target space for
static solutions of phantom EMD, and show that the null geodesics of
this target space lead, for certain discrete values of the dilaton
coupling constant, to regular multicenter black hole solutions. In
Section 3 we review the sigma model for stationary Einstein-Maxwell
theory, and construct the sigma model for stationary phantom
Einstein-Maxwell (E$\ol{\text{M}}$) theory. We then consider in
Section 4 the sigma model for five-dimensional Einstein theory with
two timelike Killing vectors, which is equivalent to stationary
phantom EMD theory for the special value $\lambda = -\sqrt3$ of the
dilaton coupling constant. We apply the matrix method to generate a
phantom charged rotating black hole solution from the Kerr solution.
In Section 5, we first review and derive the properties of the
cosets for Einstein-Maxwell-dilaton-axion (EMDA) and its phantom
counterpart (E$\ol{\text{M}}$DA), and again apply the sigma-model
approach to generate a charged rotating black hole solution of
E$\ol{\text{M}}$DA. The last section contains our conclusions.


\setcounter{equation}{0}
\section{Static phantom EMD: multicenter solutions}

We consider the string-inspired action
\begin{equation}\label{1e1}
S=-\int \dd ^{4}x \sqrt{-g}\,\;[\mathcal{R}-2\eta_1 g^{\mu\nu}
\partial_{\mu}\phi\partial_{\nu}\phi +\eta_2 \e^{2\lambda\phi}
F_{\mu\nu}F^{\mu\nu}]\,,
\end{equation}
which describes Einstein-Maxwell-dilaton gravity, with $\lambda$ the
real dilaton-Maxwell coupling constant, and $\eta_1$, $\eta_2$ the
dilaton-gravity and Maxwell-gravity coupling constants. Normal EMD
corresponds to $\eta_2=\eta_1=+1$, while phantom couplings of the
Maxwell field $F = \dd A$ or/and dilaton field $\phi$ are obtained
for $\eta_2=-1$ or/and $\eta_1=-1$. For short, we call the
corresponding theories, which include E$\ol{\text{M}}$D
($\eta_1=+1,\eta_2=-1$), EM$\ol{\text{D}}$ ($\eta_1=-1,\eta_2=+1$)
and E$\ol{\text{M}}$$\ol{\text{D}}$ ($\eta_1=-1,\eta_2=-1$), phantom
EMD.

EMD and its phantom derivatives are invariant under an
electric-magnetic duality exchanging electric and magnetic fields
and simultaneously reversing the sign of the dilaton field. For this
reason, we can restrict the investigation of static solutions to the
purely electric case, and parametrize the spacetime metric and the
electric field by
 \be
\dd s^2 = f\,\dd t^2 - f^{-1}h_{ij}\,\dd x^i\,\dd x^j, \quad F_{i0}
= \frac1{\sqrt2}\,\partial_i v
 \ee
($i=1,2,3$). Then, the original four-dimensional EMD equations can
be reduced to a three-dimensional problem deriving from the
gravitating sigma-model action \cite{GaGaKe95}
\begin{equation}\label{p3}
S_3=\int \dd^3x\sqrt{h}\big[R(h) - G_{AB}(\pmb
X)\partial_iX^A\partial_jX^Bh^{ij}\big]\,,
\end{equation}
where $R(h)$ is the Ricci scalar constructed from the 3-dimensional
metric $h_{ij}$, and $G_{AB}(\pmb X)$ is the target space metric,
with $X^1 = f$, $X^2 = v$, $X^3 = \phi$,
 \be\lb{tar} \dd l^2 =
G_{AB}\dd X^A\dd X^B = \frac{\dd f^2}{2f^2} -
\frac{\eta_2}f\,\e^{2\lambda\phi}\,\dd v^2 + 2\eta_1 \,\dd\phi^2\,.
\ee The field equations derived from~\eqref{p3} are
\begin{align}
\label{p5}& R_{ij}(h) = G_{AB}\partial_iX^A\partial_jX^B\,,\\
\label{p6}& \partial_i\left(\sqrt{h}h^{ij}G_{AB}\partial_j
X^B\right)=
 \frac{1}{2}\, \partial_A G_{BC}\partial_i X^B\partial_j X^C h^{ij}\sqrt{h}\,,
\end{align}
($\partial_A \equiv \partial /\partial X^A$). If now we assume that
the coordinates $\pmb X \equiv \pmb X(\sigma)$ depend on a single
potential function $\sigma(x^i)$, we have the freedom to choose the
potential $\sigma$ to be harmonic ($\nabla_h^2\sigma =
0$)~\cite{book}, reducing thus~\eqref{p5},~\eqref{p6} to
\begin{align}
\label{p7}& R_{ij}(h) = \frac{\dd l^2}{\dd \sigma^2}\,
\partial_i\sigma\partial_j\sigma\,,\\
\label{p8}& \frac{\dd}{\dd \sigma}\left(G_{AB}\,\frac{\dd X^B}{\dd
\sigma}\right)=
 \frac{1}{2}\, \partial_A G_{BC}\,\frac{\dd X^B}{\dd \sigma}
\frac{\dd X^C}{\dd \sigma}\,.
\end{align}
Eqs. (\ref{p8}) are the geodesic equations for the target space (\ref{tar}).
Null geodesics
\begin{equation}\label{ngeo}
G_{AB}\dot{X}^A\dot{X}^B = 0\quad (\dd l^2 = 0)
\end{equation}
(with $\dot{} \equiv \dd/\dd\sigma$) lead to a Ricci-flat, hence
flat, reduced 3-space of metric $h_{ij}$ \cite{spat,bps}. In that
case, the Laplacian $\nabla_h^2$ becomes a linear operator, so that
an arbitrary number of harmonic functions may be superposed, leading
to a multicenter solution \be\lb{multi} \sigma(\vec{r}) =
\sum_i\frac{c_i}{|\vec{r}-\vec{r_i}|} \ee (up to an additive
constant).

The null geodesic condition (\ref{ngeo}) can be interpreted as the
condition for the balance between attractive and repulsive forces
acting on the ``particles'' located at the centers
$\vec{r}=\vec{r_i}$. Defining the mass $M$, electric charge $Q$ and
dilatonic charge $D$ of an individual particle from the asymptotic
behavior (in the gauge $\phi(\infty)=0$)
 \be
f \sim 1 - \frac{2M}r \,, \quad v \sim \sqrt2\frac{Q}r \,, \quad
\phi \sim \frac{D}r\,, \quad (r \to \infty)\,,
 \ee
with $\sigma = c/r$, we see that the null geodesic condition
(\ref{ngeo}) for the metric (\ref{tar}) translates into the
condition
 \be\lb{bal}
M^2 - \eta_2Q^2 + \eta_1D^2 = 0\,.
 \ee
In the case of EMD (which generalizes the Einstein-Maxwell case),
the repulsive electrostatic force balances the attractive
gravitational and dilatonic forces. However balance can also be
achieved in the phantom cases of EM$\ol{\text{D}}$ (repulsive
electrostatic and dilatonic forces) or
E$\ol{\text{M}}$$\ol{\text{D}}$ (repulsive dilatonic force, the
electrostatic force being attractive), but not of E$\ol{\text{M}}$D
(all the forces are attractive).

The geodesics for the metric (\ref{tar}) are obtained by solving the
coupled system of equations \ba
f\ddot{f} - \dot{f}^2 & = & \eta_2f\e^{2\lambda\phi}\dot{v}^2\,, \lb{geof}\\
(f^{-1}\e^{2\lambda\phi}\dot{v})\,\dot{} & = & 0\,, \lb{geov}\\
2f\ddot{\phi} & = &
-\eta_1\eta_2\lambda\e^{2\lambda\phi}\dot{v}^2\,. \lb{geophi} \ea To
separate this system, we introduce the auxiliary function
\begin{equation}\label{p2}
\omega = \frac12\ln{f} - \lambda\phi \,.
\end{equation}
Equation (\ref{geov}) may be integrated to
\be
\dot{v}=\sqrt{2}q\e^{2\omega}\,,
\ee
with $q$ an integration constant. Replacing this in the two other geodesic
equations leads to the decoupled system
 \ba
\ddot{\phi} &=& -\eta_1\eta_2\lambda q^2\e^{2\omega}\,,\lb{ddphi}\\
\ddot{\omega}  &=& \eta_2\lambda_+q^2\e^{2\omega} \,, \lb{ddom}
 \ea
with
 \be
\lambda_+ \equiv 1+\eta_1\lambda^2\,.
 \ee
An obvious first integral of this system is
 \be\lb{k}
\lambda\dot{\omega} + \eta_1\lambda_+\dot{\phi} = k\,,
 \ee
with $k$ a second integration constant. Eq. (\ref{ddom}) may also be
first integrated to \be \dot{\omega}^2 = \eta_2\lambda_+
q^2\e^{2\omega} + a^2\,, \lb{a2} \ee with $a^2$ a third integration
constant. Finally, the null geodesic equation gives
 \be\lb{ngeo1}
(\dot{\omega}+\lambda\dot{\phi})^2 - \eta_2 q^2\e^{2\omega} +
\eta_1\dot{\phi}^2 = 0\,.
 \ee
In the generic case $\lambda_+ \neq 0$, inserting (\ref{k}) and
(\ref{a2}) into (\ref{ngeo1}) leads to the constraint between the
integration constants \be\lb{ak} a^2 + \eta_1k^2 = 0\,. \ee In the
special case  $\lambda_+ = 0$, meaning $\eta_1 = -1$ and $\lambda^2
= 1$, the comparison of (\ref{k}) and (\ref{a2}) shows that the
constraint (\ref{ak}) must again be satisfied.

To analyze the null geodesic solutions, we shall make use of the
results of \cite{phantom}. The static, spherically symmetric metric
ansatz used in \cite{phantom} is
 \be\lb{anph}
\dd s^2 = \e^{2(\omega+\lambda\phi)}\,\dd t^2 -
\e^{-2(\omega+\lambda\phi)}\left( \e^{4J}\,\dd u^2 +
\e^{2J}\,\dd\Omega^2\right)\,.
 \ee
The reduced spatial metric in (\ref{anph}) is Euclidean, $\dd s_3^2
= \dd r^2 + r^2\,\dd\Omega^2$, if $u$ is harmonic, $u = r^{-1}$, and
if $J = -\ln|u|$, corresponding to $b=0$ in Eq. (2.18) of
\cite{phantom}. Our master equation (\ref{a2}) thus coincides with
the master equation (2.14) of \cite{phantom} with the variable $u
\equiv \sigma$, while our constraint (\ref{ak}) coincides with the
constraint (2.22) of \cite{phantom} with $b=0$. We shall only
discuss here the various solutions which lead to regular
(multi)-black holes. Generically, the various centers $\vec{r} =
\vec{r_i}$ of (\ref{multi}) will correspond to horizons $\sigma \to
\pm\infty$ (however there may be other horizons, see below),
spacelike infinity  $r \to \infty$ corresponding to $\sigma = 0$. We
also expect that the solutions will be generically singular unless
all the $c_i$ in (\ref{multi}) are of the same sign, which we will
assume positive, thus restricting the domain of $\sigma$ to
$\sigma\ge0$ (this is opposite to the sign convention used in
\cite{phantom}). The solutions can be classified according to the
range of $a^2$ and the sign of $\ol{q}=\eta_2\lambda_+q^2$.
\begin{enumerate}

\item $a^2>0$, $\ol{q}<0$.
\begin{equation}\label{p14}
\e^{-2\omega}=\frac{|\ol{q}|}{a^2}\,\cosh^2[a(\sigma_0-\sigma)]\,,
\end{equation}
with $\sigma_0$ an integration constant. The metric function $f$ is
related to this by the relation (valid whenever $\lambda_+ \neq 0$)
\be\lb{fom} f = c\left(\e^{-2\omega}\e^{-2\eta_1\lambda
k\sigma}\right)^{-1/\lambda_+}\,. \ee Here $a$ is real, so $\eta_1 =
-1$ and $k = \pm |a|$ from (\ref{ak}). We find that near a horizon,
\be\lb{cosh} f \propto \exp[-2|a|\sigma/(1\mp\lambda)]\,, \ee which
is a nonanalytic function of $r$ ($\sigma = r^{-1}$). So we discard
this singular solution.

\item $a^2>0$, $\ol{q}=0$.
\begin{equation}\label{p15}
\omega=a(\sigma_0-\sigma)\,,
\end{equation}
This can be divided into two subcases. For $\lambda_+ \neq 0$
($q^2=0$), $f$ is given exactly by (\ref{cosh}) (with $|a|$ replaced
by $a$), leading again to a singular horizon. For $\lambda_+ = 0$ ,
we find
 \be
\ln f = \frac{\eta_2 q^2}{2a^2}\e^{2a(\sigma_0-\sigma)} - a\sigma +
\mbox{\rm const.}\,,
 \ee
also leading to a singular horizon.

\item $a^2>0$, $\ol{q}>0$.
\begin{equation}\label{p16}
\e^{-2\omega}=\frac{\ol{q}}{a^2}\,\sinh^2[a(\sigma_0-\sigma)]\,.
\end{equation}
Now, besides the singular horizon at $\sigma\to\pm\infty$, there is
also (if $\sigma_0 > 0$) a horizon at $\sigma = \sigma_0$ if
$\lambda_+ < 0$. The metric can be extended \`a la Kruskal across
this horizon if the metric function (\ref{fom}) is an analytic
function of $\sigma_0-\sigma$, which is the case only if $\lambda_+
= -2/p$ ($p$ positive integer). The values of the coupling constants
allowing such regular multicenter solutions are thus
 \be\lb{reg1}
\lambda^2 = \frac{p+2}p\,, \quad \eta_1 = -1\,, \quad \eta_2 = -1\,,
 \ee
corresponding to E$\ol{\text{M}}$$\ol{\text{D}}$. The explicit
solution is
 \ba\lb{sinh}
\dd s^2 &=& \left[\frac{\sinh
[a(\sigma_0-\sigma)]\e^{\pm|\lambda|a\sigma}} {\sinh
a\sigma_0}\right]^p\dd t^2 - \left[\frac{\sinh
[a(\sigma_0-\sigma)]\e^{\pm|\lambda|a\sigma}}
{\sinh a\sigma_0}\right]^{-p}\dd\vec{x}^2 \,, \nn\\
A_0 &=& \epsilon\sqrt{\frac{p}2}\,\frac{\sinh a\sigma}{\sinh
[a(\sigma_0-\sigma)]}\,,\quad \phi =
\frac{1}{2\lambda}\left((p+2)\ln\left[\frac{\sinh
[a(\sigma_0-\sigma)]}{\sinh a \sigma_0}\right] \pm
p|\lambda|a\sigma\right)\,,
 \ea
where $\epsilon=\pm1$, and we have normalized $\sigma$ by
 \be\lb{norsig}
\sum_i c_i=1\,.
 \ee
The total mass, electric charge and dilatonic charge
 \be
M = \frac{pa}2(\coth a\sigma_0\mp|\lambda|)\,, \quad Q =
\epsilon\sqrt{\frac{p}2}\,\frac{a}{\sinh a \sigma_0}\,, \quad D = -
\frac{pa}2\mbox{\rm sign}(\lambda)(|\lambda|\coth a\sigma_0\mp1)
 \ee
are related by the balance condition
 \be\lb{MQ/D}
M^2 + Q^2 = D^2\,.
 \ee
Note that in this case the singular centers $\vec{r} = \vec{r_i}$
($\sigma \to +\infty$) are hidden behind the horizon (of order $p$)
$\sigma = \sigma_0$, and that the number of connected components of
this horizon may be smaller than the number of the centers. As
discussed in Subsection IV.C, case (3.b) of \cite{phantom}, where
the Penrose diagrams for the one-center case are given, for the up
sign in (\ref{sinh}) the central singularities are spacelike for $p$
odd or timelike for $p$ even, while for the down sign in
(\ref{sinh}) they are null. The area of each horizon component is
infinite, so that these multicenter black holes have a vanishing
temperature, i.e. are cold black holes \cite{cold1,cold2}.

\item $a^2=0$, $\ol{q}>0$.
\begin{equation}\label{p17}
\e^{-2\omega}=\ol{q}\;(\sigma_0-\sigma)^2\,,
\end{equation}
leading to
 \be
f \propto |\sigma_0-\sigma|^{-2/\lambda_+}\,.
 \ee
We must distinguish between three subcases according to the value of
$\sigma_0$:

$\alpha$) $\sigma_0 > 0$. As in the preceding case, $\sigma =
\sigma_0$ is a regular horizon of order $p$ if $\lambda_+ = -2/p$
(implying $\eta_1 < 0$ and $\eta_2 < 0$ so that this case
corresponds again to E$\ol{\text{M}}$$\ol{\text{D}}$), i.e. for the
values of the coupling constants given in (\ref{reg1}). The explicit
solution is
 \ba\lb{a0al}
\dd s^2 &=& \left(1 - \frac{2M}p\sigma\right)^{p}\dd t^2 - \left(1
- \frac{2M}p\sigma\right)^{-p}\dd\vec{x}^2 \,, \nn\\
A_0 &=& \epsilon\sqrt{\frac2p}\,\frac{M\sigma}{1 -
2M\sigma/p}\,,\quad \phi =  \frac{p+2}{2\lambda}\ln\left(1 -
\frac{2M}p\sigma\right)\,,
 \ea
with $\sigma$ again normalized by (\ref{norsig}), and
$\sigma_0=p/{(2M)}$. The total mass $M$, electric charge
$Q=\epsilon(2/p)^{1/2}M$ and dilatonic charge $D=-\lambda M$ are
again related by the balance condition (\ref{MQ/D}), and the horizon
area is again infinite. The global structure in the one-center case
is discussed in Subsection IV.D, case (2) of \cite{phantom}.

$\beta$) $\sigma_0 < 0$. Then the various centers $\vec{r} =
\vec{r_i}$ each correspond to a connected component of the event
horizon $\sigma \to \infty$, which is regular of order $p$ if
$\lambda_+ = 2/p$, implying $\eta_2 > 0$. This can occur either for
$\eta_1 = +1$, i.e. for normal EMD (the corresponding multicenter
solutions have been discussed in \cite{normult,CL}), or for $\eta_1
= -1$ (EM$\ol{\text{D}}$), i.e. if
 \be\lb{reg2}
\lambda^2 = \frac{p-2}p\,, \quad \eta_1 = -1\,, \quad \eta_2 = +1\,,
 \ee
with $p \ge 2$. The explicit solution is
 \ba\lb{a0be}
\dd s^2 &=& \left(1 + \frac{2M}p\sigma\right)^{-p}\dd t^2 - \left(1
+ \frac{2M}p\sigma\right)^{p}\dd\vec{x}^2 \,, \nn\\
A_0 &=& \epsilon\sqrt{\frac2p}\frac{M\sigma}{1 + 2M\sigma/p}\,,\quad
\phi = - \frac{p-2}{2\lambda}\ln\left(1 + \frac{2M}p\sigma\right)\,.
 \ea
with $\sigma$ normalized by (\ref{norsig}), and $\sigma_0=-p/{(2M)}$.
The total electric charge and dilatonic charge are again
$Q=\epsilon(2/p)^{1/2}M$ and $D=-\lambda M$, but the balance
condition now reads
 \be
M^2 = Q^2+D^2\,.
 \ee
For $p = 2$ ($\lambda = 0$), we recover the Majumdar-Papapetrou
multi-black holes of Einstein-Maxwell theory. For $p>2$, the horizon
area is again infinite. The global structure in the one-center case
is discussed in Subsection IV.D, case (1) of \cite{phantom}.

$\gamma$) $\sigma_0 = 0$. Again the horizon $\sigma = \sigma_0$ is
regular for the values of the coupling constants (\ref{reg2}),
however $f \propto \sigma^{-2/\lambda_+}$ does not go to a constant
at spacelike infinity ($\sigma\to0$), so that the metric is
nonasymptotically flat (NAF). While regular NAF black hole solutions
of EMD are known to exist in the normal case \cite{CL}, it is easy
to show that in the present case the Ricci scalar diverges at
spatial infinity, so that these solutions do not correspond to
regular black holes\footnote{This point was overlooked in
\cite{phantom}.}.

\item $a^2<0$, $\ol{q}>0$.
In this case, $\eta_1 > 0$ from (\ref{ak}), so that $\lambda_+$ is
positive definite, and $\ol{q}>0$ implies $\eta_2>0$, corresponding
to normal (non-phantom) EMD \cite{normult,CL}.

\end{enumerate}

\setcounter{equation}{0}
\section{Stationary E$\ol{\text{M}}$ theory}\label{embm}

Now we generalize our investigations to the case of stationary
phantom EMD, i.e. we drop the restrictive assumption of staticity.
As shown in \cite{GaGaKe95}, the target spaces for stationary normal
EMD reduced to three dimensions are symmetric spaces in only two
cases: $\lambda = 0$, and $\lambda = -\sqrt3$. The third case
$\lambda = -1$ also leads to a symmetric space after enlarging the
theory to EMDA \cite{udual}. In the following we shall consider the
corresponding phantom cases, which lead to symmetric spaces only for
$\eta_1 = +1$ (which will allow us to simplify the notation and
write $\eta$ instead of $\eta_2$). We first discuss the case
$\lambda = 0$.

Let us first briefly recall the standard procedure \cite{book} for
reducing the four-dimensional Einstein-Maxwell theory deriving from
the action
\begin{equation}\label{1em}
S = -\int \dd ^{4}x \sqrt{-g}\,[\mathcal{R} +
F_{\mu\nu}F^{\mu\nu}]\,,
\end{equation}
to an effective three-dimensional gravitating sigma-model theory.
Assuming the existence of a timelike Killing vector $\partial_t$,
the metric may be parametrized by
 \be\lb{1m}
\text{d}s^2 =
f(\text{d}t - \omega_i \text{d}x^i)^2 - f^{-1}h_{ij}\,\text{d}x^i
\text{d}x^j\,,
 \ee
where $f,\,\omega_i$, and the reduced spatial metric $h_{ij}$ depend
only on the three space coordinates $x^i$ ($i=1,2,3$). The
stationary Maxwell tensor field may be split into electric and
magnetic components deriving from two three-dimensional scalar potentials $v$
(electric) and $u$ (magnetic) according to
 \be\lb{1vu}
F_{i0} = \partial_i v\,, \qquad F^{ij} = fh^{-1/2}\epsilon^{ijk}
\partial_k u\,,
 \ee
($\epsilon^{ijk}$ is the totally antisymmetric symbol, with
$\epsilon^{123}=1$). Finally,
solving the mixed $G_0^i = 0$ Einstein equations enables to trade
the three-dimensional vector $\omega_i$ for the three-dimensional
scalar twist potential $\chi$ such that
 \be\lb{1twist}
\partial_i\chi = -f^2h^{-1/2}h_{ij}\epsilon^{jkl}\partial_k\omega_l
+2(u\partial_i v - v\partial_i u)\,.
 \ee
The remaining Einstein-Maxwell equations then reduce to the
following three-dimensional Ernst equations~\cite{er}
 \ba\lb{1ernst1}
f\nabla^2{\cal E} & = & \nabla{\cal E} \cdot (\nabla{\cal E} +
2\ol{\psi}\nabla\psi)\,,\nonumber \\
f\nabla^2\psi & = & \nabla\psi \cdot (\nabla{\cal E} +
2\ol{\psi}\nabla\psi)\,, \\
f^2R_{ij}(h) & = & {\rm Re} \left[\frac{1}{2}\,{\cal
E},_{(i}\ol{{\cal E}},_{j)} + 2\psi{\cal E},_{(i}\ol{\psi},_{j)}
-2{\cal E}\psi,_{(i}\ol{\psi},_{j)} \right], \nonumber
 \ea
where the complex Ernst potentials are defined by
\begin{equation}\label{1ernst2}
    \mathcal{E} = f + \text{i} \chi -\ol{\psi}\psi\,, \quad \psi = v +
    \text{i}u\,,
\end{equation}
and the dot product and covariant derivative $\nabla$ are defined in
terms of the spatial metric $h_{ij}$. The gravitating sigma-model
equations (\ref{1ernst1}) are invariant under an eight-parameter
group $SU(2,1)$ of transformations acting on the target space
coordinates $\E$ and $\psi$ and leaving invariant the spatial metric
$h_{ij}$. The action of this group may be linearized by introducing
the complex Kinnersley potentials ~\cite{kin} ($U,V,W$), related to
the Ernst potentials by
\begin{equation}\label{1ernst3}
    {\cal E} =\frac{U-W}{U+W}\,,\quad \psi = \frac{V}{U+W}\,.
\end{equation}
This parametrization is only apparently redundant, the Kinnersley
potentials being defined by (\ref{1ernst3}) up to a complex local
rescaling $\Phi(\vec{x}) \to \xi(\vec{x})\Phi(\vec{x})$
($\xi(\vec{x})\in C$) of the ``spinor'' $\Phi = (U,V,W)$, allowing
e.g. to fix the value of the norm $\parallel\Phi\parallel^2 = |U|^2
+ |V|^2 -|W|^2$. The gravistatic potential $f$ is simply related to
the Kinnersley potentials by
 \be f = {\rm Re}\,\E + \ol{\psi}\psi = \frac{|U|^2 +
|V|^2 -|W|^2}{|U+W|^2} \,.
 \ee

Now, consider the general action
\begin{equation}\label{1f}
S = -\int \dd ^{4}x \sqrt{-g}\,[\mathcal{R}+\eta
F_{\mu\nu}F^{\mu\nu}]\,,
\end{equation}
which describes the Einstein-Maxwell theory if $\eta=+1$, and the
Einstein-anti-Maxwell theory if $\eta=-1$. The reduction of the
stationary Einstein-anti-Maxwell theory to an effective
three-dimensional gravitating sigma model closely parallels that of
the Einstein-Maxwell theory, so that it is sufficient to identify
the equations where the sign of $\eta$ (or, in other words, the sign
of the gravitational constant) comes in. The definition
\eqref{1twist} of the twist potential becomes
 \be\lb{1twist2}
\partial_i\chi =
-f^2h^{-1/2}h_{ij}\epsilon^{jkl}\partial_k\omega_l
+2\eta(u\partial_i v - v\partial_i u)\,,
 \ee
and the definition of the generalized Ernst potentials
 \be\lb{1ernst4}
\mathcal{E} = f + \text{i} \chi -\eta\ol{\psi}\psi\,,\quad \psi = v
+ \text{i}u\,,
 \ee
leads to the generalized Ernst equations
 \ba\lb{1ernst5}
f\nabla^2{\cal E} & = & \nabla{\cal E} \cdot (\nabla{\cal E} +
2\eta\ol{\psi}\nabla\psi)\,,\nonumber \\
f\nabla^2\psi & = & \nabla\psi \cdot (\nabla{\cal E} +
2\eta\ol{\psi}\nabla\psi)\,, \\
f^2R_{ij}(h) & = & {\rm Re} \left[\frac{1}{2}\,{\cal
E},_{(i}\ol{{\cal E}},_{j)} + 2\eta\psi{\cal E},_{(i}\ol{\psi},_{j)}
-2\eta{\cal E}\psi,_{(i}\ol{\psi},_{j)} \right]. \nonumber
 \ea
These equations may also be obtained from the equations
(\ref{1ernst1}) by replacing $\E$ with $\eta\E$ and $f$ with  $\eta f$,
suggesting the
definition of the generalized Kinnersley potentials ($U,V,W$)
\begin{equation}\label{1ernst6}
    {\cal E} =\eta\,\frac{U-W}{U+W}\,,\quad \psi = \frac{V}{U+W}\,,
\end{equation}
in terms of which we express the scalar function $f$ as
\begin{equation}\label{1ernst7}
    f= \eta\,\frac{|U|^2+|V|^2-|W|^2}{|U+W|^2}\,.
\end{equation}
Using~\eqref{1ernst6} and~\eqref{1ernst7} in
equations~\eqref{1ernst5}, the latter, once written in terms of
($U,V,W$), take the same form as equations~\eqref{1ernst1} do when
they are expressed in terms of the previously defined Kinnersley
potentials in~\eqref{1ernst3}. Thus, equations~\eqref{1ernst5}
remain invariant when acted upon by $SU(2,1)$, which keeps the norm
$|U|^2+|V|^2-|W|^2$ invariant so that the sign of $f$ is also
invariant. Then, using the third line in~\eqref{1ernst5} one
concludes that the spatial metric $h_{ij}$ is also invariant.

Asymptotically flat field configurations are such that, up to a
gauge transformation, $f(\infty)=1$ with the three other scalar
potentials $\chi$, $v$ and $u$ vanishing at spatial infinity,
leading, from the third equation (\ref{1ernst5}), to an
asymptotically flat reduced spatial metric $h_{ij}$. Thus, at
spatial infinity ($|\vec{x}|\to\infty$), the spinor
$\Phi_{\eta}(\infty)$ representing any asymptotically flat solution
is of the form
 \ba
\label{1s3}& \Phi_+(\infty) = (1,0,0) & \qquad (\eta = +1)\,, \\
\label{1s4}& \Phi_-(\infty) = (0,0,1) & \qquad (\eta = -1)\,.
 \ea
The asymptotic behavior (\ref{1s3}) is preserved, up to a complex
rescaling, by the transformations of the subgroup $U(1,1)$ mixing
$V$ and $W$ on the one hand, by the transformations of the subgroup
$U(1)$ changing the phase of $U$ on the other hand, and more
generally by the product of two such transformations up to a complex
rescaling, i.e. by the transformations of the isotropy subgroup $H_+
= S[U(1,1)\times U(1)] \subset SU(2,1).$ The coset space for
stationary Einstein-Maxwell theory ($\eta= +1$) is thus
$SU(2,1)/S[U(1,1)\times U(1)]$ \cite{BMG}. Similarly, the asymptotic
behavior (\ref{1s4}) is preserved (up to a rescaling) by the $U(2)$
transformations mixing $U$ and $V$ on the one hand, and by the
$U(1)$ transformations acting on $W$ on the other hand, generating
the isotropy subgroup $H_- = S[U(2)\times U(1)].$ So the coset space
for stationary Einstein-anti-Maxwell theory ($\eta= -1$) is
$SU(2,1)/S[U(2)\times U(1)]$.

A simple application is the generation of static phantom charged
black holes from the Schwarzschild solution $f_0 = 1-2M_0/r$. The
spinor representing the Schwarzschild solution is (up to a
rescaling)
 \be
\Phi_0 = (M_0,0,r-M_0)\,.
 \ee
Acting on this by $SO(2)$ transformations parametrized by an angle
$\alpha$ leads to \be \Phi = (M_0\cos\alpha,M_0\sin\alpha,r-M_0)\,,
\ee which corresponds to the phantom Reissner-Nordstr\"om solution
\cite{gira}
 \be
\dd s^2 =
\left(1-\frac{r_+}{\bar{r}}\right)\left(1-\frac{r_-}{\bar{r}}\right)\,
\dd t^2 -
\left(1-\frac{r_+}{\bar{r}}\right)^{-1}\left(1-\frac{r_-}{\bar{r}}
\right)^{-1}\dd\bar{r}^2 - \bar{r}^2\,\dd\Omega^2\,, \quad \psi =
\frac{Q}{\bar{r}}\,, \ee with $\bar{r} = r + r_-$, $r_+ =
M_0(1+\cos\alpha)$, $r_- = -M_0(1-\cos\alpha)$, $Q = M_0\sin\alpha$.
The black hole mass $M = (r_+ + r_-)/2 = M_0\cos\alpha$ vanishes for
$\alpha=\pi/2$ \footnote{The corresponding $SO(2)$ transformation is
$U = -V_0, V = U_0$. In the EM case, this transformation generates
from the Schwarzschild solution the non-asymptotically flat
Bertotti-Robinson solution \cite {kerr}, which is not possible
here.}. Then the black hole mass becomes negative, until the
limiting value $M=-M_0$ is attained, corresponding (not
surprisingly) to the singular Schwarzschild solution with negative
mass.


\setcounter{equation}{0}
\section{Stationary phantom Kaluza-Klein theory}\label{ss}

Let us consider the vacuum Einstein-Hilbert action in 5D
\begin{equation}\label{1}
S_{5}=-\int \dd ^{5}x \sqrt{-^{(5)}g}\;\; ^{(5)}\mathcal{R}\, .
\end{equation}
The theory may be reduced \`a la Kaluza-Klein to four dimensions by
assuming the existence of a Killing vector  $\partial_{5}$ which may
be either spacelike (as usually assumed), corresponding to
$\eta=+1$, or timelike (this is the phantom case), corresponding to
$\eta=-1$. The generalized Kaluza-Klein ansatz
 \be\lb{kk}
\dd s^{2}_{(5)} =\e^{2\phi/\sqrt{3}\,}\;
^{(4)}g_{\mu\nu}(x^{\alpha})\, \dd x^{\mu}\dd x^{\nu} -
\eta\e^{-4\phi/\sqrt{3}}(\dd x^{5}+2A_{\mu}\, \dd x^{\mu})^{2}\,,
 \ee
leads to the reduced action
\begin{equation}\label{2}
S_{4} = -\int\dd ^{4}x \sqrt{-^{(4)}g}\,\;[ ^{(4)}\mathcal{R}-2\;
^{(4)}g^{\mu\nu}\partial_{\mu}\phi\partial_{\nu}\phi +
\eta\e^{-2\sqrt{3}\phi} F_{\mu\nu}F^{\mu\nu}]\, ,
\end{equation}
for normal ($\eta=+1$), or phantom ($\eta=-1$)
Einstein-Maxwell-dilaton (EMD or E$\ol{\text{M}}$D) theory
\cite{phantom} with dilaton coupling constant $\lambda = -\sqrt3$.

The four-dimensional theory (\ref{2}) may further be reduced to
three dimensions by assuming stationarity, i.e. the existence of a
second, timelike, Killing vector $\partial_{0}$. Equivalently, we
can directly reduce the five-dimensional theory (\ref{1}) to three
dimensions by assuming the existence of two commuting Killing
vectors $\partial_{0}$ and $\partial_{5}$, with the first timelike
and the second spacelike or timelike according to the sign of
$\eta$. We require the reduced three-space to be topologically
Euclidean, so that the signature of the five-dimensional spacetime
metric is $+ - - - -$ for $\eta = +1$, and $+ + - - -$ for
$\eta=-1$\footnote{The phantom ($\eta=-1$) Kaluza-Klein theory is a
two-time theory. For this reason, we do not require the fifth
dimension to be compact, which would lead unavoidably to closed
timelike curves.}. The five-to-three reduction is parametrized
according to the metric ansatz \cite{ms}
 \ba\lb{5to3}
ds_{(5)}^2 &=&\lambda_{ab}[\dd x^{a}+V^{a}{}_{i}\,\dd x^{i}][\dd
x^{b} + V^{b}{}_{j}\,\dd x^{j}] -\,\tau^{-1}h_{ij}\,\dd x^{i}\dd
x^{j}\; ,
\end{eqnarray}
where $a,b=0,5$ and $\tau =\left|\det{[\lambda_{ab}]}\right|$. The 3-vectors
$V^{a}{}_{i}$ may be dualized to scalar twist potentials $\omega_a$ via the
relations
\begin{equation}\label{dual}
\omega_{a,i}=\left|h\right|^{-1/2}\tau
h_{il}\lambda_{ab}\epsilon^{jkl}V^{b}{}_{j,k}\,.
\end{equation}
The remaining field equations derived from~\eqref{1} may be written
as the three-dimensional sigma-model system \cite{ms}
\begin{eqnarray}
& & R_{ij}(h)=\dfrac{1}{4}\,\tr[\chi^{-1}\partial_{i}\chi \chi^{-1}
\partial_{j}\chi ]\label{7a} \,,\\
& & \nabla_{i}\left(\chi^{-1}\partial^{i}\chi
\right)=0\label{7b} \,,
\end{eqnarray}
where $\chi$ is the $3\times3$ symmetric and unimodular Maison matrix defined
(in block form) by
\begin{equation}\label{6}
\chi = \begin{bmatrix}
\lambda + \tau^{-1}\omega\omega^T & \tau^{-1}\omega \\
\tau^{-1}\omega^T & \tau^{-1}
\end{bmatrix}\,.
\end{equation}

The field equations (\ref{7a}), (\ref{7b}) are invariant under an
$SL(3,\mathbb{R})$ group of transformations $P$ acting bilinearly on
the matrix $\chi$:
 \be\lb{transf}
\chi' = P^T\chi P\,.
 \ee
For asymptotically flat configurations ($\lambda(\infty) =$
diag$(1,-\eta)$, $\omega(\infty) = 0$), the matrix $\chi$ is
asymptotic to
\begin{equation}\label{9a}
\chi(\infty) =\begin{bmatrix}
1&0&0\\
0&-\eta &0\\
0&0&1
\end{bmatrix}\,.
\end{equation}
It follows that the isotropy group at spatial infinity is $H_+ =
SO(2,1)$ for $\eta=+1$, and $H_- = SO(3)$ for $\eta=-1$. Thus, the
coset space for phantom Kaluza-Klein theory is
$SL(3,\mathbb{R})/SO(3)$ (instead of $SL(3,\mathbb{R})/SO(2,1)$ in
the normal case).

To illustrate the applications of this sigma model, we shall
generate from the four-dimensional Kerr metric embedded in five
dimensions (with a timelike fifth dimension),
 \be\label{kerr}
\dd s_{(5)}^{2} =\dfrac{g_0}{\Sigma_{0}}\,(\dd t-\Omega_{0}\,\dd
\varphi)^{2} + (\dd x^5)^2
-\dfrac{\Sigma_{0}}{g_0}\bigg[\dfrac{g_0}{\Delta_{0}}\,\dd r^{2} +
g_0\,\dd \theta^{2}+\Delta_{0}\sin^{2}\theta \,\dd
\varphi^{2}\bigg]\,,
 \ee
where
 \ba
&\Delta_{0}=r^{2}-2M_0r+a_{0}^{2}\,, & \Sigma_{0}=r^{2}
+a_{0}^{2}\cos^{2}\theta\,,\lb{DS}\\
& g_0=\Sigma_0-2M_or\,, &
\Omega_{0}=-2a_{0}M_0\dfrac{r\sin^{2}\theta}{g_0}\,, \lb{gO}
\end{eqnarray}
a charged rotating black hole solution of the phantom theory
E$\ol{\text{M}}$D. The metric (\ref{kerr}) is of the form (\ref{5to3})
with ${V^0}_3=-\Omega_0$. Using the duality equation (\ref{dual}),
we obtain
 \be\lb{om0}
\omega_0 = 2a_0M_0\frac{\cos\theta}{\Sigma_0} \,,
 \ee
leading to the seed Maison matrix
\begin{eqnarray}\label{23}
\chi_0  = g_0^{-1}\begin{bmatrix}
(r-2M_0)^{2}+a_{0}^{2}\cos^{2}\theta &0&2a_{0}M_0\cos\theta \\
0&g_0&0\\
2a_{0}M_0\cos\theta &0&\Sigma_{0}
\end{bmatrix}.
\end{eqnarray}
Applying to this seed the $SO(3)$ group transformation
\begin{equation}\label{alkk}
\chi = P^{T}\chi_0P\,, \quad P=\begin{bmatrix}
\cos\alpha &-\sin\alpha &0\\
\sin\alpha &\cos\alpha &0\\
0&0&1
\end{bmatrix}\,,
\end{equation}
we obtain the Maison matrix for the charged solution
\begin{equation}\label{24}
\chi(r)=\frac{1}{g_0}\begin{bmatrix}
g_0-2M_0\cos^{2}\alpha (r-2M_0) &2M_0\sin\alpha\cos\alpha (r-2M_0)&2a_{0}M_0\cos\alpha\cos\theta \\
2M_0\sin\alpha\cos\alpha (r-2M_0)&g_0-2M_0\sin^{2}\alpha (r-2M_0)&-2a_{0}M_0\sin\alpha\cos\theta\\
2a_{0}M_0\cos\alpha\cos\theta &-2a_{0}M_0\sin\alpha\cos\theta
&\Sigma_{0}
\end{bmatrix}.
\end{equation}
After inverse dualization, this leads to the 5-metric
 \be\label{27}
\dd s_{(5)}^{2} =
\dfrac{\Sigma}{\Sigma_0}\,\Lambda^{2}+\dfrac{4Q}{\Sigma_0}\,r\Lambda\Psi
+ \dfrac{g}{\Sigma_0}\,\Psi^{2}-\dfrac{\Sigma_0}{g_0}
\bigg[\dfrac{g_0}{\Delta_0}\,\dd r^{2}+g_0\,\dd \theta^{2}+\Delta_0\sin^{2}\theta \,\dd \varphi^{2}\bigg]\,,
 \ee
where
 \ba
& \Sigma=\Sigma_0-4\sigma r\,, & g = g_0 + 4\sigma r\,, \\
& \Lambda = \dd
x^{5}-\dfrac{2JQ}{(M-\sigma)}\dfrac{r\sin^{2}\theta}{g_0}\,\dd
\varphi\,, & \Psi =\dd t+2J\,\dfrac{r\sin^{2}\theta}{g_0}\,\dd
\varphi\,,
\end{eqnarray}
and
\begin{eqnarray}\label{26}
& &M=M_0(1+\cos^{2}\alpha)/2\; ,\;\;Q=M_0\sin\alpha\cos\alpha\,,\nonumber\\
& &J=a_{0}M_0\cos\alpha\; ,\;\;\sigma =M_0\sin^{2}\alpha/2\,,
\end{eqnarray}
are the mass, electric charge, angular momentum and rescaled
dilatonic charge (the usual dilatonic charge is $D =
\sqrt3\,\sigma$) defined from the asymptotic behavior of the charged
solution
 \be\label{25}
\chi(r\rightarrow +\infty)  \sim
\begin{bmatrix}
1-\dfrac{2(M-\sigma)}{r} & \dfrac{2Q}{r} & \dfrac{2J\cos\theta}{r^{2}} \\
\dfrac{2Q}{r} & 1-\dfrac{4\sigma}{r} & 0\\
\dfrac{2J\cos\theta}{r^{2}} & 0 & 1+\dfrac{2(M+\sigma)}{r}
\end{bmatrix}.
 \ee
Note that only three of these charges are independent, as
$Q^2 = 2\sigma(M-\sigma)$, with $0 \le \sigma \le M$. The seed Kerr
parameters may be recovered from these by
 \be
M_0 = M+\sigma\,, \quad a_0 = \frac{|J|}{\sqrt{M^2-\sigma^2}}\,.
 \ee
The solution (\ref{27}) is
the phantom version of Rasheed's rotating solution~\cite{ra} with
vanishing magnetic charge ($P=0$), from which it may be obtained by
making the substitution $Q^{2}\rightarrow -Q^{2}$ and changing the
sign of the dilatonic charge.

The Kaluza-Klein reduction (\ref{kk}) of (\ref{27}) leads to the
four-dimensional solution
\begin{eqnarray}
\dd s_{(4)}^{2}&=&\dfrac{g_0}{\sqrt{\Sigma_0\Sigma}}\bigg[\dd
t+2J\,\dfrac{r\sin^{2}\theta}{g_0}\dd \varphi\bigg]^2 -
\dfrac{\sqrt{\Sigma_0\Sigma}}{g_0}\bigg[\dfrac{g_0}{\Delta_0}\,\dd
r^{2}+
g_0\,\dd \theta^{2}+\Delta_0\sin^{2}\theta \,\dd \varphi^{2}\bigg]\,,\lb{28a}\\
A&=&Q\dfrac{r}{\Sigma} \left[\,\dd
t-\dfrac{J}{M-\sigma}\sin^{2}\theta \,\dd \varphi\right]\,, \quad
\e^{2\phi/\sqrt{3}}=\sqrt{\dfrac{\Sigma_0}{\Sigma}}\,.\lb{28b}
 \ea
which describes an asymptotically flat rotating phantom dilatonic
black hole with electric charge $Q$ and magnetic dipole moment
 \be\lb{dip}
\mu = \frac{JQ}{M-\sigma}=a_0M_0\sin\alpha\,.
 \ee
The corresponding gyromagnetic ratio
 \be
g = \frac{2M\mu}{QJ} = 2 + \tan^2\alpha
 \ee
is for $\alpha \neq 0$ larger than $2$. In the static ($a_0 = 0$)
case, the solution (\ref{28a})-(\ref{28b}) reduces to the ``cosh'' solution of
\cite{phantom} with $\lambda=-\sqrt3$,
\begin{eqnarray}\label{14}
\dd s_{(4)}^{2}&=&\left(1-\frac{r_+}{\bar{r}}\right)
\left(1-\frac{r_-}{\bar{r}}\right)^{-1/2}\,
\dd t^{2}-\left(1-\frac{r_+}{\bar{r}}\right)^{-1}
\left(1-\frac{r_-}{\bar{r}}\right)^{1/2}[\dd \bar{r}^{2}+
(\bar{r}-r_+)(\bar{r}-r_-)\,\dd \Omega^2]\,,\nn\\
A_{0}&=&\dfrac{Q}{\bar{r}}\,, \qquad
\e^{4\phi/\sqrt{3}}=1-\frac{r_-}{\bar{r}}\,,
\end{eqnarray}
where $\bar{r} = r - 4\sigma$, and $r_+ = 2(M-\sigma) \ge 0$,
$r_-=-4\sigma \le 0$.

The five-dimensional solution (\ref{27}) or its Kaluza-Klein
reduction (\ref{28a}) represent a black hole with the same horizons
 \be\lb{hork}
r = r_{h\pm} \equiv M_0 \pm \sqrt{M_0^2-a_0^2}
 \ee
(if $a_0^2 \le M_0^2$) and ring singularity $r = 0$, $\cos\theta =
0$ as the seed Kerr black hole. The four-dimensional metric
(\ref{28a}) is also singular on the surfaces $\Sigma(r,\theta) = 0$,
 \be\lb{singkk}
r = r_{s\pm}(\theta) \equiv 2\sigma \pm
(4\sigma^2-a_0^2\cos^2\theta)^{1/2}\,.
 \ee
The largest value of $r_{s+}(\theta)$ is $4\sigma$, so this
singularity will be naked unless
 \be
4\sigma < M_0 + \sqrt{M_0^2-a_0^2}\,, \quad \mbox{\rm or} \quad
\sqrt{M_0^2-a_0^2} > -M_0\cos(2\alpha)\,.
 \ee
This regularity condition is satisfied if either $\cos(2\alpha) >
0$, or if $\cos(2\alpha) < 0$ and $|a_0| < 2|Q|$. However at the
five-dimensional level (metric (\ref{27})), $\Sigma = 0$ is simply
the static limit associated with the time coordinate $x^5$, and is a
mere coordinate singularity. So, in the five-dimensional setting,
the four-dimensional singularity (\ref{singkk}) is actually a
spurious singularity due to the Kaluza-Klein reduction.

When the charging parameter $\alpha$ takes the value $\alpha =
\pi/2$, something curious happens. For this value, the physical
charges are
 \be
M = \sigma\,, \quad Q = 0\,, \quad J = 0
 \ee
(implying $\Sigma = g_0$ and $g=\Sigma_0$), so that at first sight the
solution is electrically neutral and static. Actually it is neither, as
the dipole magnetic moment $\mu$ given by (\ref{dip}) goes to the finite
value $a_0M_0$ for $\alpha \to \pi/2$, showing that the static ($J=0$)
four-dimensional metric (\ref{28a}) is generated by the magnetic dipole field
 \be
A_3 = -a_0M_0\,r\sin^2\theta/\Sigma\,.
 \ee
Again, this is just an effect of the Kaluza-Klein reduction. The
five-dimensional metric (\ref{27}) reads, for $\alpha = \pi/2$,
 \be
\dd s_{(5)}^{2} = \dd t^2 + \dfrac{g_0}{\Sigma_{0}}\,(\dd x^5 +
\Omega_{0}\,\dd \varphi)^{2}
-\dfrac{\Sigma_{0}}{g_0}\bigg[\dfrac{g_0}{\Delta_{0}}\,\dd r^{2} +
g_0\,\dd \theta^{2}+\Delta_{0}\sin^{2}\theta \,\dd
\varphi^{2}\bigg]\,,
 \ee
which is just the embedded Kerr metric (\ref{kerr}) with the time
coordinates $t=x^0$ and $x^5$ exchanged and the angular momentum
flipped. For the intermediate value $\alpha = \pi/4$ (corresponding
to the maximum electric charge $Q=M-\sigma=M_0/2$), we find $\Sigma
= g$, so that the reduction of (\ref{27}) can be indifferently
carried out relative to $\partial_0$ or to $\partial_{5}$, leading
to the same reduced four-dimensional fields, provided the sign of
$\varphi$ is flipped. For other values of $\alpha$, the exchange
$\alpha = \pi/2-\alpha'$, $\varphi = -\varphi'$ is equivalent to the
exchange $(\Sigma,g) = (g',\Sigma')$ between the two time
coordinates in (\ref{27}), leading to the exchange between the
effective four-dimensional charges
\begin{eqnarray}
&M'=\dfrac{M+3\sigma}2\,,\quad &\sigma' =\frac{M-\sigma}2\,,\nonumber\\
&J'=\dfrac{JQ}{M-\sigma}\,,\quad &Q'=Q\,.
\end{eqnarray}

The reason for this becomes clear if we realize that the
$SL(3,\mathbb{R})$ group of transformations (\ref{transf}) includes
a $GL(2,R)$ subgroup of transformations
 \be
P = \begin{bmatrix} \Pi & 0 \\ 0 & -|\Pi|^{-1} \end{bmatrix}
 \ee
where the $2\times2$ matrix $\Pi$ defines a coordinate transformation in the
space of the two timelike cyclic coordinates,
 \be
\begin{bmatrix}x^{0} \\ x^5 \end{bmatrix} = \Pi
\begin{bmatrix}x^{'0} \\ x^{'5} \end{bmatrix}\,.
 \ee
The charging transformation (\ref{alkk}) is such a $GL(2,R)$
transformation. This means that the five-dimensional ``charged''
black hole metric (\ref{27}) is simply the five-dimensional
(phantom) Kerr black  string metric (\ref{kerr}) transformed to an
unfamiliar coordinate system,  while the four-dimensional  charged
black hole solution (\ref{28a})-(\ref{28b}) is simply the Kaluza-Klein reduction
of the  phantom Kerr black string relative to a linear combination
of the Killing vectors $\partial_0$ and $\partial_5$.

\setcounter{equation}{0}
\section{Stationary E$\ol{\text{M}}$DA theory}\label{emda}

Einstein-Maxwell-dilaton-axion theory is a truncation of the bosonic
sector of four-dimensional {\cal N} = 4 supergravity, defined by the
action
\begin{equation}\label{5.1}
    S=-\int \dd x^4\sqrt{-g}\big[{\cal R}
-2\partial_{\mu}\phi\partial^{\mu}\phi-
    \frac{1}{2}\,\e^{4\phi}\partial_{\mu}\kappa\partial^{\mu}\kappa
    +\eta\e^{-2\phi}F_{\mu\nu}F^{\mu\nu}+
    \kappa F_{\mu\nu}\tilde{F}^{\mu\nu}\big]\,,
\end{equation}
where $\kappa$ is the pseudoscalar axion, and ${\tilde
F}^{\mu\nu}=\frac{1}{2}\,E^{\mu\nu\lambda\tau}F_{\lambda\tau\,}$
\footnote{We use the convention $E^{\mu\nu\lambda\tau} \equiv
|g|^{-1/2}\varepsilon^{\mu\nu\lambda\tau}$, with $\varepsilon^{0123}
= -1$.}. The sign $\eta=+1$, corresponding to normal EMDA, can be
changed to $\eta=-1$, corresponding to phantom E$\ol{\text{M}}$DA,
by the formal translation $\phi\rightarrow \phi + \ii \pi/2$.
Reduction of the stationary theory to three dimensions is achieved
in a manner similar to that of Einstein-Maxwell theory, with a
metric parametrized as in~\eqref{1m}, and electric ($v$),
magnetic($u$), and twist ($\chi$) potentials defined by
\begin{align}
\label{5e3} & F_{i0} = \frac{1}{\sqrt{2}}\,\partial_i v\,, \quad
\kappa \tilde{F}^{ij} + \eta \e^{-2\phi}F^{ij} =
\frac{f}{\sqrt{2}}h^ {-1/2}\,\epsilon^{ijk}\partial_k u\, ,\\
\label{5e4} & \partial_i\chi +v\partial_i u-u\partial_i v =
-f^2h^{-1/2}h_{ij}\epsilon^{jkl}\partial_k\omega_l\,.
\end{align}
As shown in \cite{udual} (where we replace whenever necessary $\e^{2\phi}$
by $\eta\e^{2\phi}$), this reduction leads to the symmetric target
space metric
\begin{align}
\dd l^2
= & \frac12\,f^{-2}\dd f^2 + \frac12\,f^{-2}(\dd \chi + v\dd u - u\dd
v)^2 - \eta f^{-1}\e^{-2\phi}\dd v^2 \nn\\
\label{5.5} & - \eta f^{-1}\e^{2\phi}(\dd u - \kappa \dd v)^2 +
2\dd \phi^2 + \frac12\,\e^{4\phi}\dd \kappa^2\,,
\end{align}
with signature $(++++--)$ in the normal case $\eta = +1$ and $(++++++)$
in the phantom case $\eta = -1$. This target space is invariant under
a ten-parameter group $Sp(4,\mathbb{R})$ of transformations \cite{udual}.

The symplectic group $Sp(4,\mathbb{R})$ is the group of real $4\times 4$
matrices $M$ satisfying
\be\label{5.6}
M^TJM = J\,,\quad J = \begin{bmatrix} 0 & \sigma_0 \\
-\sigma_0 & 0 \end{bmatrix}\,,
\ee
where $\sigma_0$ is the $2\times 2$ identity matrix. A base of ten traceless
$4\times 4$ matrices generating the algebra $sp(4,\mathbb{R})$ has been
constructed in~\cite{udual}:
\begin{eqnarray}
\label{5e1} V_a=\frac{1}{2}\begin{bmatrix}
0&\sigma_a\\
\sigma_a &0
\end{bmatrix}\;\;  ,\;\; W_a=\frac{1}{2}
\begin{bmatrix}
\sigma_a & 0\\
0&-\sigma_a
\end{bmatrix}\,,\\
\label{5e2} U_a=\frac{1}{2}
\begin{bmatrix}
0&\sigma_a\\
-\sigma_a &0
\end{bmatrix}\;\; ,\;\; U_2=\frac{1}{2}
\begin{bmatrix}
\sigma_2 &0\\
0&\sigma_2
\end{bmatrix}\label{}\,,
\end{eqnarray}
($a = 0,1,3$, $\sigma_1 = \sigma_x$, $\sigma_2 = i\sigma_y$,
$\sigma_3 = \sigma_z$ with $\sigma_x,\,\sigma_y,\,\sigma_z$ the
Pauli matrices). The six generators $V_a$ and $W_a$ are symmetrical,
and the four generators $U_a$ and $U_2$ are antisymmetrical.

As also shown in \cite{udual}, the reduced field equations derive
from the gravitating sigma-model action
\begin{equation}\label{5.3}
S_3 = \int \dd ^3x\sqrt{h}\Big[R(h) + \frac{1}{4}\,\tr (\nabla M
\nabla M^{-1})\Big]\,,
\end{equation}
with the symmetrical symplectic matrix representative \be\lb{mare}
M=
\begin{bmatrix}
P^{-1}&P^{-1}Q\\
QP^{-1}&P+QP^{-1}Q
\end{bmatrix},
\ee where the $2\times2$ block matrices $P$ and $Q$ are \be\lb{PQ}
P= \e^{-2\phi}\begin{bmatrix}
f\e^{2\phi}-\eta v^2& -\eta v\\
-\eta v&-\eta
\end{bmatrix}, \;
Q=\begin{bmatrix}
vw -\chi & w\\
w &-\kappa
\end{bmatrix}\qquad (w=u-\kappa v).
\ee

The matrix representative $M$ parametrizes an element of the coset
space $Sp(4,\mathbb{R})/H_{\eta}\,$, where $H_{\eta}$ is the
isotropy subgroup that preserves asymptotic behavior. To determine
this subgroup, we consider asymptotically flat solutions for which
$f(\infty)=1$ and the five other target space coordinates go to
zero. For normal EMDA ($\eta=+1$), we find from (\ref{mare}) and
(\ref{PQ}) that
 \be
M_+(\infty) = \begin{bmatrix} \sigma_3 & 0 \\ 0 & \sigma_3
\end{bmatrix} \,.
 \ee
This remains invariant under the $Sp(4,\mathbb{R})$ transformations
 \be\lb{Sp4}
M' = P^TMP
 \ee
generated by
\begin{equation}\label{5.13}
    {\rm lie}\,(H_+)=(V_1,W_1,U_0,U_3)\,.
\end{equation}
These matrices generate the $u(1,1)$ algebra with center $U_3$:
$[W_1,V_1] = U_0\,$, $[U_0,V_1] =W_1\,$, $[W_1,U_0] =V_1$ and
$[V_1,U_3]=[W_1,U_3]=[U_0,U_3]=0$. It follows that the coset for
EMDA is $Sp(4,\mathbb{R})/U(1,1) \sim SO(3,2)/(SO(2,1)\times
SO(2))$.

For phantom EMDA ($\eta=-1$), $M_-(\infty)$ is the $4\times 4$
identity matrix,
 \be M_-(\infty) = \begin{bmatrix} \sigma_0 & 0 \\ 0
& \sigma_0
\end{bmatrix} \,.
 \ee
This matrix remains invariant under the subgroup generated by the
antisymmetrical matrices
\begin{equation}\label{5.12}
    {\rm lie}\,(H_-)=(U_a,U_2)\,.
\end{equation}
These matrices generate the $u(2)$ algebra with center $U_0$:
$[U_i,U_j] = - \epsilon_{ijk}U_k\,$, $[U_0,U_i] = 0$, $i,j,k =
1,2,3$ and $\epsilon_{123}=+1$. It follows that the coset for
E$\ol{\text{M}}$DA is $Sp(4,\mathbb{R})/U(2)\sim
SO(3,2)/(SO(3)\times SO(2))$.

A third four-dimensional subgroup of $Sp(4,\mathbb{R})$ is
$GL(2,\mathbb{R})$. The choice of this as isotropy subgroup leads to
the coset $Sp(4,\mathbb{R})/GL(2,\mathbb{R}) \sim
SO(3,2)/(SO(2,1)\times SO(1,1))$~\cite{Gilmore}, which is the coset
for Euclidean EMDA~\cite{eemda}.

Let us now, as in the previous section, apply the E$\ol{\text{M}}$DA
sigma model to the generation of a charged rotating black hole
solution from the four-dimensional Kerr metric given by (\ref{kerr})
(without the fifth dimension). The twist potential $\chi_0$ is the
opposite of $\omega_0$ in (\ref{om0}), leading to the seed matrix
representative
 \be
M_0 = g_0^{-1}\begin{bmatrix}
\Sigma_0 & 0 & 2a_0M_0\cos\theta & 0\\
0 & g_0 & 0 & 0\\
2a_0M_0\cos\theta & 0 & (r-2M_0)^2 + a_0^2\cos^2\theta & 0\\
0 & 0 & 0 & g_0
\end{bmatrix}.
\end{equation}
Applying to this the $Sp(4,\mathbb{R})$ transformation (\ref{Sp4}) with
\begin{equation}
P=\begin{bmatrix}
\cos\alpha & -\sin\alpha & 0 & 0\\
\sin\alpha & \cos\alpha & 0 & 0\\
0 & 0 & \cos\alpha & -\sin\alpha\\
0 & 0 & \sin\alpha & \cos\alpha
\end{bmatrix}\,,
\end{equation}
we obtain the matrix representing the charged rotating solution
 \be
M = g_0^{-1}\begin{bmatrix} r(r-2D) + a_0^2c^2 & -\sqrt2Qr & 2a_0Mc
& -a_0\sqrt2Qc \\
-\sqrt2Qr & r(r-2M) + a_0^2c^2 & -a_0\sqrt2Qc & 2a_0Dc \\
2a_0Mc & -a_0\sqrt2Qc & (r-2M_0)(r-2M) + a_0^2c^2 & 2M_0(r-2M_0) \\
-a_0\sqrt2Qc & 2a_0Dc & 2M_0(r-2M_0) & (r-2M_0)(r-2D) + a_0^2c^2
\end{bmatrix},\label{chemda}
 \ee
with $c \equiv \cos\theta$, and
 \be\lb{MQD}
M=M_0\cos^2\alpha\; ,\;\; Q=\sqrt2M_0\sin\alpha\cos\alpha\; ,\;\;
D=M_0\sin^2\alpha \quad (M_0=M+D\,, \;\; Q^2 = 2MD)\,.
 \ee
The corresponding fields, obtained by identifying (\ref{mare}) with
(\ref{chemda}) and performing inverse dualization, are
 \ba\lb{kemda}
\dd s^{2}&=&\dfrac{g_0}{\Sigma}\bigg[\dd
t+2J\,\dfrac{r\sin^{2}\theta}{g_0}\dd \varphi\bigg]^2 -
\dfrac{\Sigma}{g_0}\bigg[\dfrac{g_0}{\Delta_0}\,\dd r^{2}+
g_0\,\dd \theta^{2}+\Delta_0\sin^{2}\theta \,\dd \varphi^{2}\bigg]\,,\nn\\
A&=&Q\dfrac{r}{\Sigma} \left[\,\dd t-a_0\sin^{2}\theta \,\dd
\varphi\right]\,, \quad \e^{2\phi} = \dfrac{\Sigma_0}{\Sigma}\,,
\quad \kappa = - \frac{2a_0D\cos\theta}{\Sigma_0}\,,
 \ea
with
 \be
\Sigma= r(r-2D)+a_0^2\cos^2\theta\,, \quad J = a_0M\,.
 \ee
In the static ($a_0=0$) case, this solution reduces to the cosh
solution of \cite{phantom} with $\lambda=-1$,
\begin{eqnarray}
\dd s^{2}&=&\left(1-\frac{r_+}{\bar{r}}\right) \, \dd
t^{2}-\left(1-\frac{r_+}{\bar{r}}\right)^{-1} [\dd \bar{r}^{2}+
(\bar{r}-r_+)(\bar{r}-r_-)\,\dd \Omega^2]\,,\nn\\
A_{0}&=&\dfrac{Q}{\bar{r}}\,, \qquad
\e^{2\phi}=1-\frac{r_-}{\bar{r}}\,,
\end{eqnarray}
with $r_+ = 2M \ge 0$, $r_-=-2D\le 0$, and $\bar{r} = r+r_-$.

The solution (\ref{kemda}) of phantom EMDA represents a rotating,
electrically charged black hole with mass $M$, electric charge $Q$,
dilatonic charge $D$, angular momentum $J$, magnetic dipole charge
$a_0Q$, and axionic dipole charge $a_0D$. It is formally identical
to the charged rotating black hole solution of normal EMDA
\cite{sen}, the only difference being that the dilatonic charge $D$
is negative in the case of EMDA ($Q^2 = -2MD$) and positive in the
present case of E$\ol{\text{M}}$DA ($Q^2 = 2MD$). A consequence is
that (as in the case of the phantom $\lambda = -\sqrt3$ black hole
(\ref{28a})) the outermost singular surface ($\Sigma(r,\theta) = 0$),
 \be
r_{s+}(\theta) = D + \sqrt{D^2-a_0^2\cos^2\theta}
 \ee
may lie (in part) outside the event horizon (\ref{hork}). The
largest value of $r_{s+}(\theta)$ is $2D$, so this singularity will
be naked unless
 \be
2D < M_0 + \sqrt{M_0^2-a_0^2} \,, \quad \mbox{\rm or} \quad
\sqrt{M_0^2-a_0^2} > -M_0\cos(2\alpha)\,.
 \ee
This is always satisfied if $\cos(2\alpha) > 0$ ($M>D$). In that
case, the spacetime of metric (\ref{kemda}) is a regular black hole
for $|a_0| \le M_0 = M+D$, extremality being achieved for $|a_0| =
M+D$. If $\cos(2\alpha) < 0$ ($M<D$), the condition for black hole
regularity is stronger:
 \be
|a_0| < \sqrt2|Q|\,.
 \ee
In particular, the value $\alpha=\pi/2$ leads from (\ref{MQD}) to
the massless, neutral solution
 \ba
\dd s^{2}&=& \dd t - \dfrac{g_0}{\Delta_0}\,\dd r^{2}-
g_0\,\dd \theta^{2}-\Delta_0\sin^{2}\theta \,\dd \varphi^{2}\,,\nn\\
A&=& 0 \,, \quad \e^{2\phi}= 1 + \dfrac{2Dr}{g_0}\,, \quad \kappa =
- \frac{2a_0D\cos\theta}{\Sigma_0}\,,
 \ea
with a static metric but non-vanishing axionic dipole charge $a_0D$.
This solution is singular ($g_0 = \Sigma$ vanishes for $r =
r_{s+}(\theta) > r_{h+}$), including in the static ($a_0=0$) case.


\setcounter{equation}{0}
\section{Conclusion}

In this paper, we first constructed static multicenter solutions of
phantom EMD. We found that these lead to regular black holes in two
cases. The first case is that of E$\ol{\text{M}}$$\ol{\text{D}}$
(phantom Maxwell and dilaton fields), for the discrete values
$\lambda^2 = (p+2)/p$ ($p$ positive integer) of the dilaton coupling
constant. In this case the singular centers are hidden behind the
horizon of order $p$. The number of connected components of this
horizon may be equal to, or smaller than, the number $n$ (unrelated
to $p$) of the centers. In particular, the integration constants can
be arranged so that the solution represents a black hole without
spatial symmetry, with a single horizon component hiding the $n$
singularities. The second case corresponds to EM$\ol{\text{D}}$
(phantom dilaton field), for the discrete values $\lambda^2 =
(p-2)/p$ ($p>2$ integer). In this case, which generalizes the
Majumdar-Papapetrou multi-black holes of Einstein-Maxwell theory,
each center corresponds to a connected component of the event
horizon. All these multicenter black holes have an infinite horizon
area and vanishing temperature.

We then considered sigma models for the special values of the
dilatonic coupling constant $\lambda=0$, $\lambda=-1$ and
$\lambda=-\sqrt3$, for which stationary phantom EMD (or EMDA in the
case $\lambda=-1
$) reduced to three dimensions has a symmetric
target space $G/H$. This occurs only for a non-phantom (or
vanishing) dilaton field, so that these sigma models do not admit
multicenter solutions. In all cases, the noncompact isotropy
subgroup $H_+$ corresponds to the normal case and the compact
isotropy subgroup $H_-$ corresponds to the phantom case. This can be
understood if one notes that changing the sign of the gravitational
coupling constant $\eta_2$ of the Maxwell field is equivalent to
changing the sign of the spacetime metric. Thus reduction of the
four-dimensional phantom theory relative to a timelike Killing
vector is equivalent to reduction of the normal theory relative to a
spacelike Killing vector, which is well-known \cite{BMG} to lead
to a compact isotropy subgroup.

We also in each case applied a particular sigma-model transformation
to generate phantom static and rotating charged solutions from a
neutral solution. While the charging transformation in the normal
case belongs to an $SO(1,1)$ subgroup of $H_+$, it belongs in the
phantom case to an $SO(2)$ subgroup of $H_-$, resulting in important
differences between the two cases. For the normal theories, an
infinite $SO(1,1)$ boost leads, in the static case, to an extreme
black hole saturating the lower bound $M^2 \ge Q^2/(1+\lambda^2)$
\cite{normult}. For the phantom theories there is, for $\lambda^2
\le 1$, no such lower bound and massless solutions are possible
\cite{gira}, as obtained here in the cases of E$\ol{\text{M}}$ and
E$\ol{\text{M}}$DA for an $SO(2)$ rotation of $\pi/2$. Moreover, the
extreme solution, obtained by a rotation of $\pi$ in the case of
E$\ol{\text{M}}$, or by a rotation of $\pi/2$ in the cases of
phantom Kaluza-Klein theory or of E$\ol{\text{M}}$DA, is singular.

We have seen that, in the case of phantom Kaluza-Klein theory, the
naked singularity of this static extreme solution, or more generally
that of a sector of the family of phantom rotating charged
solutions, becomes a mere coordinate singularity once the
four-dimensional solution is lifted to five dimensions. By analogy
with this, and with similar mechanisms of higher-dimensional
resolution of naked curvature singularities \cite{GHT}, one can
speculate that the naked singularities of a sector of the family of
solutions (\ref{kemda}) of E$\ol{\text{M}}$DA could be resolved by
lifting the theory to some higher-dimensional theory.

\vspace{0.5cm} \noindent {\bf Acknowledgement:} M.E.R. and J.C.F.
thank CNPq (Brazil) for partial financial support.

\end{document}